\documentclass{ifacconf}

\usepackage{graphicx}      
\usepackage{natbib}        

\usepackage{color}
\def\JB{\textcolor{blue}}

\usepackage{textcomp}
\usepackage{url}
\usepackage{algorithm}
\usepackage{comment}
\usepackage{amsfonts,amssymb,amsmath,mathtools}
\usepackage{latexsym}
\newtheorem{theorem}{\bf Theorem} \newtheorem{definition}[theorem]{\bf Definition} 
\newtheorem{lemma}[theorem]{\bf Lemma} 
   
\newtheorem{assumption}[theorem]{\bf Assumption}  
\newtheorem{Algorithm}[theorem]{\bf Algorithm}

\begin{document}
\begin{frontmatter}

\title{Data-Driven Tracking MPC for Changing Setpoints} 

\thanks[footnoteinfo]{This work was funded by Deutsche Forschungsgemeinschaft (DFG,
German Research Foundation) under Germany’s Excellence Strategy - EXC
2075 - 390740016. The authors thank the International Max Planck Research
School for Intelligent Systems (IMPRS-IS) for supporting Julian Berberich,
and the International Research Training Group Soft Tissue Robotics (GRK
2198/1).\\
\textcopyright 2020 the authors. This work has been accepted to IFAC for publication under a 
Creative Commons Licence CC-BY-NC-ND.}

\author[First]{Julian Berberich} 
\author[First]{Johannes K\"ohler} 
\author[Second]{Matthias A. M\"uller}
\author[First]{Frank Allg\"ower}

\address[First]{Institute for Systems Theory and Automatic Control, University of Stuttgart, 70550 Stuttgart, Germany (email:$\{$ julian.berberich, johannes.koehler, frank.allgower$\}$@ist.uni-stuttgart.de).}
\address[Second]{Leibniz University Hannover, Institute of Automatic Control, 30167 Hannover, Germany (e-mail:mueller@irt.uni-hannover.de).}

\begin{abstract}
We propose a data-driven tracking model predictive control (MPC) scheme to control unknown discrete-time linear time-invariant systems.
The scheme uses a purely data-driven system parametrization to predict future trajectories based on behavioral systems theory.
The control objective is tracking of a given input-output setpoint.
We prove that this setpoint is exponentially stable for the closed loop of the proposed MPC, if it is 
reachable by the system dynamics and constraints.
For an unreachable setpoint, our scheme guarantees closed-loop exponential stability of the optimal reachable equilibrium.
Moreover, in case the system dynamics are known, the presented results extend the existing results for model-based setpoint tracking to the case where the stage cost is only positive semidefinite in the state.
The effectiveness of the proposed approach is illustrated by means of a practical example.
\end{abstract}


\end{frontmatter}

\section{Introduction}
In the behavioral approach to control, it was proven by~\cite{Willems05} that a single, persistently exciting trajectory of a linear time-invariant (LTI) system suffices to reproduce all system trajectories, without explicitly identifying the system.
Recently, various contributions have exploited this result to develop data-driven analysis and control methods in the classical state-space control framework.
For instance, the work of~\cite{Willems05} is used for data-driven dissipativity verification in~\cite{Romer19} and it is extended to certain classes of nonlinear systems in~\cite{Berberich20a}.
Further, state- and output-feedback controllers from noise-free data are designed in~\cite{Persis19}, whereas~\cite{Berberich20b} consider robust controller design from noisy data.
Finally,~\cite{Waarde19} investigate data-driven control problems without requiring persistently exciting data.
These works illustrate a potential advantage of direct data-driven methods compared to identification-based controller design, leading to simple implementations with desirable end-to-end-guarantees.
Similarly, data-driven model predictive control (MPC) schemes based on~\cite{Willems05} have been suggested in~\cite{Yang15} as well as~\cite{Coulson19,Coulson19b}.
A first theoretical analysis of such an MPC scheme is provided in~\cite{Berberich19b}, which utilizes terminal equality constraints to prove closed-loop exponential stability, also for the case of output measurement noise.
The present paper extends the approach of~\cite{Berberich19b} to tracking of setpoints, which may be unreachable and change online, by introducing an artificial equilibrium to the MPC scheme which is optimized online.
Compared to the MPC scheme of~\cite{Berberich19b}, this has the additional advantages of improved robustness, a larger region of attraction, and the possibility to consider setpoints, which are not feasible equilibria for the \emph{unknown} system.
The basic idea of optimizing an artificial setpoint online was introduced in the model-based setting in~\cite{limon2008mpc}, and extended, e.g., to nonlinear systems in~\cite{limon2018nonlinear} or to dynamic reference trajectories in~\cite{Koehler19b}.
In this paper, we prove closed-loop recursive feasibility, constraint satisfaction, as well as exponential stability of the desired equilibrium, if it is reachable by the dynamics and constraints, and of the optimal reachable equilibrium otherwise.
Our work provides a solution to the data-driven tracking problem, which has only recently been addressed under significantly more restrictive conditions on the data, cf.~\cite{Salvador19}.
Further, our main technical contribution is also an extension of the model-based tracking MPC approach of~\cite{limon2008mpc}, where so far only positive definite stage costs have been considered.

The paper is structured as follows.
After introducing some preliminaries in Section~\ref{sec:setting}, we state the data-driven tracking MPC scheme in Section~\ref{sec:MPC}, and we prove the desired properties of the closed loop.
The applicability of the presented method is illustrated via an example in Section~\ref{sec:example}, and the paper is concluded in Section~\ref{sec:conclusion}.

\section{Preliminaries}\label{sec:setting}
\subsection{Notation}

$\mathbb{I}_{[a,b]}$ denotes the set of integers in the interval $[a,b]$.
For a vector $x$ and a symmetric matrix $P=P^\top\succeq0$, we write $\lVert x\rVert_P=\sqrt{x^\top Px}$.
Further, we denote the minimal and maximal eigenvalue of P by $\lambda_{\min}(P)$ and $\lambda_{\max}(P)$,
respectively. 
For multiple matrices $P_i$, $i=1,\dots,n$, we write $\lambda_{\min}(P_1,\dots,P_n)=\min\{\lambda_{\min}(P_1),\dots,\lambda_{\min}(P_n)\}$, and similarly for the maximal eigenvalue.
Moreover, $\lVert x\rVert_2$, $\lVert x\rVert_1$, and $\lVert x\rVert_\infty$ denote the standard Euclidean, $\ell_1$-, and $\ell_\infty$-norm of $x\in\mathbb{R}^n$, respectively.
If the argument is matrix-valued, then we mean the induced norm.
For a sequence $\{x_k\}_{k=0}^{N-1}$, we define the Hankel matrix
\begin{align*}
H_L&(x)\coloneqq\begin{bmatrix}x_0 & x_1& \dots & x_{N-L}\\
x_1 & x_2 & \dots & x_{N-L+1}\\
\vdots & \vdots & \ddots & \vdots\\
x_{L-1} & x_{L} & \dots & x_{N-1}
\end{bmatrix}.
\end{align*}
A stacked window of the sequence is written as
\begin{align*}
x_{[a,b]}=\begin{bmatrix}x_a\\\vdots\\x_b\end{bmatrix}.
\end{align*}
Moreover, we write $x$ either for the sequence itself or for the stacked vector $x_{[0,N-1]}$.

\subsection{Data-driven system representation}
We consider the following standard definition of persistence of excitation.

\begin{definition}
We say that a signal $\{u_k\}_{k=0}^{N-1}$ with $u_k\in\mathbb{R}^m$ is persistently exciting of order $L$ if $\text{rank}(H_L(u))=mL$.
\end{definition}

Throughout the paper, we consider discrete-time LTI systems of order $n$ with $m$ inputs and $p$ outputs.
It will be assumed that the system order $n$ is known, but all of our results remain true if $n$ is replaced by an upper bound.

\begin{definition}\label{def:trajectory_of}
We say that an input-output sequence $\{u_k,y_k\}_{k=0}^{N-1}$ is a trajectory of an LTI system $G$, if there exists an initial condition $\bar{x}\in\mathbb{R}^n$ as well as a state sequence $\{x_k\}_{k=0}^{N}$ such that
\begin{align*}
x_{k+1}&=Ax_k+Bu_k,\>\>x_0=\bar{x}\\
y_k&=Cx_k+Du_k,
\end{align*}
for $k=0,\dots,N-1$, where $(A,B,C,D)$ is a minimal realization of $G$.
\end{definition}

Throughout this paper, we assume that the input-output behavior of the unknown system can be explained by a (controllable and observable) minimal realization (cf. Definition~\ref{def:trajectory_of}).
The following is the main result of~\cite{Willems05} and will be of central importance throughout the paper.
While it is originally formulated in the behavioral approach, we employ the formulation in the state-space control framework from~\cite{Berberich20a}.

\begin{theorem}\label{thm:traj_rep}
Suppose $\{u_k^d,y_k^d\}_{k=0}^{N-1}$ is a trajectory of an LTI system $G$, where $u^d$ is persistently exciting of order $L+n$.
Then, $\{\bar{u}_k,\bar{y}_k\}_{k=0}^{L-1}$ is a trajectory of $G$ if and only if there exists $\alpha\in\mathbb{R}^{N-L+1}$ such that
\begin{align}\label{eq:thm_hankel}
\begin{bmatrix}H_L(u^d)\\H_L(y^d)\end{bmatrix}\alpha
=\begin{bmatrix}\bar{u}\\\bar{y}\end{bmatrix}.
\end{align}
\end{theorem}
Theorem~\ref{thm:traj_rep} states that the data-dependent Hankel matrices in~\eqref{eq:thm_hankel} span the trajectory space of the system $G$, given that the input data is persistently exciting.
Equivalently, one input-output trajectory can be used to reconstruct any other system trajectory, by forming linear combinations of its time-shifts.
In Section~\ref{sec:MPC}, we use Theorem~\ref{thm:traj_rep} to set up a data-driven MPC scheme and we prove desirable closed-loop properties.
In particular, we show closed-loop exponential stability of the optimal reachable input-output equilibrium, as defined in the following section.

\subsection{Input-output equilibria}
Since state measurements are not available in the present setting, we define an equilibrium in terms of an input-output pair as follows.

\begin{definition}\label{def:equil}
We say that an input-output pair $(u^s,y^s)\in\mathbb{R}^{m+p}$ is an equilibrium of an LTI system $G$, if the sequence $\{\bar{u}_k,\bar{y}_k\}_{k=0}^{\JB{n}}$ with $(\bar{u}_k,\bar{y}_k)=(u^s,y^s)$ for all $k\in\mathbb{I}_{[0,\JB{n}]}$ is a trajectory of $G$.
\end{definition}

Thus, an input-output pair $(u^s,y^s)$ is an equilibrium if, while applying $\JB{n+1}$ times the input $u^s$ to the system, the output stays constant at $y^s$.
This implies that the internal state in any minimal realization, denoted by $x^s$, stays constant.
We write $u^s_n$ and $y^s_n$ for vectors containing $n$ times the input and output component of an equilibrium, respectively.
In this paper, we use MPC to steer the system towards a target input-output setpoint, while the input and output satisfy pointwise-in-time constraints, i.e., $u_t\in\mathbb{U},y_t\in\mathbb{Y}$, for all $t\in\mathbb{N}$.
The sets $\mathbb{U}$ and $\mathbb{Y}$ are assumed to be convex.
We do not require that the target setpoint $(u^T,y^T)\in\mathbb{R}^{m+p}$ is an equilibrium of the system or satisfies the constraints, but our scheme will achieve convergence to the \emph{optimal reachable equilibrium}.
For matrices $S\succeq0,T\succ0$, the optimal reachable equilibrium $(u^{sr},y^{sr})$ is defined as the feasible equilibrium $(u^s,y^s)$, which minimizes $\lVert u^s-u^T\rVert_S^2+\lVert y^s-y^T\rVert_T^2$.
Due to a local controllability argument in the proof of our main result, only equilibria which are strictly inside the constraints can be considered, i.e., equilibria which lie in some convex set $\mathbb{U}^s\times\mathbb{Y}^s\subseteq\text{int}(\mathbb{U}\times\mathbb{Y})$.
Let $\{u^d,y^d\}_{k=0}^{N-1}$ be a measured trajectory of $G$ and suppose that $u^d$ is persistently exciting of order \JB{$2n+1$}.
Then, $(u^{sr},y^{sr})$ is the minimizer of
\begin{align}\label{eq:opt_reach_eq}
J_{eq}^*(u^T,y^T)=
\underset{u^s,y^s,\alpha}{\min}\>\>&\lVert u^s-u^T\rVert_S^2+\lVert y^s-y^T\rVert_T^2\\\nonumber
s.t.\quad&\begin{bmatrix}H_{\JB{n+1}}\left(u^d\right)\\H_{\JB{n+1}}\left(y^d\right)\end{bmatrix}\alpha=\begin{bmatrix}u_{\JB{n+1}}^s\\y_{\JB{n+1}}^s\end{bmatrix},\\\nonumber
&u^s\in\mathbb{U}^s,\>y^s\in\mathbb{Y}^s.
\end{align}
Although this is not required for the implementation of the proposed scheme, Problem~\eqref{eq:opt_reach_eq} can be used to compute $(u^{sr},y^{sr})$ from measured data.
Clearly, Problem~\eqref{eq:opt_reach_eq} is convex.
Similar to~\cite{limon2008mpc,Koehler19b}, we require that it is even strongly convex, as captured in the following assumption.

\begin{assumption}\label{ass:eq_convex}
The optimization problem~\eqref{eq:opt_reach_eq} is strongly convex w.r.t. $(u^s,y^s)$.
\end{assumption}

Assumption~\ref{ass:eq_convex} implies that the optimal reachable equilibrium $(u^{sr},y^{sr})$ is unique, and it is, e.g., satisfied if $S\succ0$.
%
More generally, in case that $S\succeq0$, it is also satisfied if $D=0$ and the matrix $B$ has full column rank in some (and hence any) minimal realization, since these two conditions imply that there exists a unique equilibrium input for any equilibrium output.
Throughout this paper, it will be assumed that $B$ has full column rank.
This assumption excludes over-actuated systems and guarantees the existence of a unique equilibrium input-output pair for any steady state.
To be more precise, it holds for any equilibrium $(u^s,y^s)$ with corresponding steady state $x^s$ that
\begin{align}\label{eq:equilibrium_unique}
(I-A)x^s=Bu^s,\quad y^s=Cx^s+Du^s.
\end{align}
Since $B$ has full column rank,~\eqref{eq:equilibrium_unique} implies
\begin{align}\label{eq:bound_xuy_2}
\lVert u^s\rVert_2^2+\lVert y^s\rVert_2^2\leq c_{x,1}\lVert x^s\rVert_2^2,
\end{align}
for $c_{x,1}=\lVert C\rVert_2^2+(1+\lVert D\rVert_2^2)\left\lVert B^\dagger(I-A)\right\rVert_2^2$, where $B^\dagger$ is the Moore-Penrose inverse of $B$.
Conversely, since we consider minimal realizations, it follows directly from the system dynamics that there exists a constant $c_{x,2}>0$ such that
\begin{align}\label{eq:bound_xuy_1}
\lVert x^s\rVert_2^2\leq c_{x,2}(\lVert u^s\rVert_2^2+\lVert y^s\rVert_2^2).
\end{align}

\section{Data-driven tracking MPC}\label{sec:MPC}

In this section, we propose a data-driven MPC scheme for setpoint tracking, which is essentially a combination of the schemes from~\cite{Berberich19b} and~\cite{limon2008mpc}.
To be more precise, the scheme relies on Theorem~\ref{thm:traj_rep} to predict future trajectories and contains a standard tracking cost as well as terminal equality constraints w.r.t. an input-output setpoint, similar to~\cite{Berberich19b}.
Moreover, in line with~\cite{limon2008mpc}, this setpoint is optimized online and its deviation from the desired target setpoint is penalized in the cost.
After stating the scheme in Section~\ref{sec:MPC_scheme}, we prove recursive feasibility, closed-loop constraint satisfaction, and exponential stability of the optimal reachable equilibrium in Section~\ref{sec:MPC_thm}.

\subsection{Tracking MPC scheme}\label{sec:MPC_scheme}
To set up a data-driven MPC scheme based on Theorem~\ref{thm:traj_rep}, we require three ingredients:
a) an initially measured input-output data trajectory $\{u_k^d,y_k^d\}_{k=0}^{N-1}$ used for prediction via Theorem~\ref{thm:traj_rep}, b) past $n$ input-output measurements $u_{[t-n,t-1]}$, $y_{[t-n,t-1]}$ to specify initial conditions, and c) a target input-output setpoint $(u^T,y^T)$.
Given these components, we define the open-loop optimal control problem
\begin{subequations}\label{eq:tracking_MPC}
\begin{align}\nonumber
J_L^*(&u_{[t-n,t-1]},y_{[t-n,t-1]},u^T,y^T)=\\\nonumber
\underset{\substack{\alpha(t),u^s(t),y^s(t)\\\bar{u}(t),\bar{y}(t)}}{\min}\>\>&\sum_{k=0}^{\JB{L}}\lVert \bar{u}_k(t)-u^s(t)\rVert_R^2+\lVert \bar{y}_k(t)-y^s(t)\rVert_Q^2\\\nonumber
&+\lVert u^s(t)-u^T\rVert_S^2+\lVert y^s(t)-y^T\rVert_T^2\\\label{eq:tracking_MPC1}
s.t.\quad&\begin{bmatrix}\bar{u}_{[-n,\JB{L}]}(t)\\\bar{y}_{[-n,\JB{L}]}(t)\end{bmatrix}=\begin{bmatrix}H_{\JB{L+n+1}}(u^d)\\H_{\JB{L+n+1}}(y^d)\end{bmatrix}\alpha(t),\\\label{eq:tracking_MPC2}
&\begin{bmatrix}\bar{u}_{[-n,-1]}(t)\\\bar{y}_{[-n,-1]}(t)\end{bmatrix}=\begin{bmatrix}u_{[t-n,t-1]}\\y_{[t-n,t-1]}\end{bmatrix},\\\label{eq:tracking_MPC3}
&\begin{bmatrix}\bar{u}_{[L-n,\JB{L}]}(t)\\\bar{y}_{[L-n,\JB{L}]}(t)\end{bmatrix}=\begin{bmatrix}u^s_{\JB{n+1}}(t)\\y^s_{\JB{n+1}}(t)\end{bmatrix},\\\label{eq:tracking_MPC4}
&\bar{u}_k(t)\in\mathbb{U},\>\>\bar{y}_k(t)\in\mathbb{Y},\>\>k\in\mathbb{I}_{[0,\JB{L}]},\\
&(u^s(t),y^s(t))\in\mathbb{U}^s\times\mathbb{Y}^s.
\end{align}
\end{subequations}

The constraint~\eqref{eq:tracking_MPC1} replaces the model and parametrizes all possible trajectories of the unknown LTI system, assuming persistence of excitation of the input $u^d$.
Moreover, the initial and terminal constraints over $n$ \JB{and $n+1$} steps in~\eqref{eq:tracking_MPC2} and~\eqref{eq:tracking_MPC3} imply that the states $\bar{x}_0(t)$ and $\bar{x}_{\JB{L}}(t)$, corresponding to the predicted input-output trajectory $\{\bar{u}_k(t),\bar{y}_k(t)\}_{k=0}^{\JB{L}}$, are equal to the internal state $x_t$ and the setpoint $x^s$, respectively, in any minimal realization.
Problem~\eqref{eq:tracking_MPC} is similar to the nominal MPC problem proposed in~\cite{Berberich19b}, with the main difference that the desired input-output setpoint is replaced by an artificial equilibrium $(u^s(t),y^s(t))$, which is optimized online.
Moreover, its distance w.r.t. $(u^T,y^T)$ is penalized in the cost.
Note that the terminal constraint~\eqref{eq:tracking_MPC3} and the dynamics~\eqref{eq:tracking_MPC1} imply that $(u^s(t),y^s(t))$ is indeed an equilibrium of the system.

Problem~\eqref{eq:tracking_MPC} requires only a single measured input-output trajectory and can thus be implemented directly, without any model knowledge.
As an advantage over the existing data-driven MPC scheme of~\cite{Berberich19b}, the target setpoint $(u^T,y^T)$ can be arbitrary and is not required to be reachable or even an equilibrium for the unknown system dynamics.
If $\mathbb{U}$ and $\mathbb{Y}$ are polytopic (quadratic), then~\eqref{eq:tracking_MPC} is a convex (quadratically constrained) quadratic program, which can be solved efficiently.
As is standard in MPC, Problem~\eqref{eq:tracking_MPC} is solved in a receding horizon fashion, compare Algorithm~\ref{alg:MPC}.

\begin{algorithm}
\begin{Algorithm}\label{alg:MPC}
\normalfont{\textbf{Data-Driven Tracking MPC Scheme}}
\begin{enumerate}
\item At time $t$, take the past $n$ measurements $u_{[t-n,t-1]},y_{[t-n,t-1]}$ and solve~\eqref{eq:tracking_MPC}.
\item Apply the input $u_t=\bar{u}_0^*(t)$.
\item Set $t=t+1$ and go back to (1).
\end{enumerate}
\end{Algorithm}
\end{algorithm}

We assume for the stage cost that $Q,R\succ0$.
The open-loop cost and the optimal open-loop cost of~\eqref{eq:tracking_MPC} are denoted by $J_L\left(x_t,u^T,y^T,u^s(t),y^s(t),\alpha(t)\right)$ and $J_L^*\left(x_t,u^T,y^T\right)$, where $x_t$ is the state at time $t$, corresponding to $(u_{[t-n,t-1]},y_{[t-n,t-1]})$ in some minimal realization.

\subsection{Closed-loop guarantees}\label{sec:MPC_thm}

In this section, we prove that the MPC scheme defined via~\eqref{eq:tracking_MPC} exponentially stabilizes the optimal reachable equilibrium state $x^{sr}$ and thus, also the input and output converge exponentially to $u^{sr}$ and $y^{sr}$, respectively.
For this, $(u^T,y^T)$ is not required to satisfy the constraints or to be an equilibrium in the sense of Definition~\ref{def:equil}, in which case $(u^{sr},y^{sr})\neq(u^T,y^T)$.
As in the model-based case, recursive feasibility of the scheme will be guaranteed, even if the target setpoint $(u^T,y^T)$ changes online.
As an additional technical contribution, the result of this section extends the (model-based) setpoint tracking MPC analysis of~\cite{limon2008mpc} to the case that the stage cost is only positive semidefinite in the state.
This is relevant in a model-based setting, e.g., if input-output models are used for prediction.

As will become clear in the proof of our main result, the fact that the stage cost may not be positive definite in the state complicates the analysis of the MPC scheme.
To overcome this issue, we analyze the closed loop of the proposed ($1$-step) MPC scheme over $n$ consecutive time steps and show a desired Lyapunov function decay over $n$ steps.
Moreover, as in~\cite{Berberich19b}, we exploit detectability of the stage cost via an input-output-to-state stability (IOSS) Lyapunov function.
For some state in an (observable) minimal realization, there exists an IOSS Lyapunov function $W(x)=\lVert x\rVert_P^2$ with $P\succ0$ which satisfies
\begin{align*}
W(Ax+Bu)-W(x)\leq -\frac{1}{2}\lVert x\rVert_2^2+c_1\lVert u\rVert_2^2+c_2\lVert y\rVert_2^2,
\end{align*}
for all $x\in\mathbb{R}^n,u\in\mathbb{R}^m,y=Cx+Du$, with some $c_1,c_2>0$, cf.~\cite{cai2008input}.
For some $\gamma>0$, we define a Lyapunov function candidate based on the IOSS Lyapunov function $W$, the optimal cost $J_L^*$, and the cost of the optimal reachable equilibrium $J^*_{eq}(u^T,y^T)$, cf.~\eqref{eq:opt_reach_eq}, as
\begin{align*}
V(x_t,u^T,y^T)&=J_L^*(x_t,u^T,y^T)+\gamma W(x_t-x^{sr})\\
&\quad-J_{eq}^*(u^T,y^T),
\end{align*}
where $x^{sr}$ is the state corresponding to $(u^{sr},y^{sr})$.
Using this Lyapunov function candidate, the following result proves recursive feasibility, constraint satisfaction, and exponential stability of the closed loop.

\begin{theorem}\label{thm:main}
Suppose that $\mathbb{U}$ and $\mathbb{Y}$ are compact, $L\geq2n$, Assumption~\ref{ass:eq_convex} holds, and $u^d$ is persistently exciting of order \JB{$L+2n+1$}.
If the MPC problem~\eqref{eq:tracking_MPC} is feasible at initial time $t=0$, then
\begin{enumerate}
\item[(a)] it is feasible at any $t\in\mathbb{N}$,
\item[(b)] the closed loop satisfies the constraints, i.e., $u_t\in\mathbb{U}$ and $y_t\in\mathbb{Y}$ for all $t\in\mathbb{N}$,
\item[(c)] the optimal reachable equilibrium $x^{sr}$ is 	exponentially stable for the resulting closed loop.
\end{enumerate}
\end{theorem}
\begin{pf}
\textbf{(a). Recursive Feasibility}\\
For the artificial equilibrium, we choose as a candidate at time $t+1$ the previously optimal one, i.e., $u^s\text{$'$}(t+1)=u^{s*}(t),y^s\text{$'$}(t+1)=y^{s*}(t)$.
Moreover, for the input-output predictions, we consider the standard candidate solution, consisting of the previously optimal solution shifted by one step and appended by $(u^s\text{$'$}(t+1),y^s\text{$'$}(t+1))$, i.e.,
\begin{align*}
\bar{u}_k'(t+1)=\bar{u}_{k+1}^*(t),\>\>\bar{y}_k'(t+1)=\bar{y}_{k+1}^*(t),\>\>k\in\mathbb{I}_{[-n,\JB{L-1}]},
\end{align*}
and $\bar{u}_{\JB{L}}'(t+1)=u^s\text{$'$}(t+1),\bar{y}_{\JB{L}}'(t+1)=y^s\text{$'$}(t+1)$.
Finally, according to Theorem~\ref{thm:traj_rep}, there exists an $\alpha'(t+1)$ satisfying~\eqref{eq:tracking_MPC1}.\\
\textbf{(b). Constraint Satisfaction}\\
This follows directly from recursive feasibility, together with Theorem~\ref{thm:traj_rep} and the constraints of~\eqref{eq:tracking_MPC}.\\
\textbf{(c). Exponential Stability}\\
We first show that the Lyapunov function candidate $V$ is quadratically lower and upper bounded.
Thereafter, we prove that $V$ is non-increasing and decreases exponentially over $n$ time steps, which implies exponential stability for the closed loop.
\\
\textbf{(c.1). Lower Bound on $\mathbf{V}$}\\
Using that $J_{eq}^*(u^T,y^T)\leq \lVert u^s-u^T\rVert_S^2+\lVert y^s-y^T\rVert_T^2$ for any equilibrium $(u^s,y^s)$ satisfying the constraints of~\eqref{eq:opt_reach_eq}, $V$ is quadratically lower bounded as
\begin{align*}
\gamma\lambda_{\min}(P)\lVert x_t-x^{sr}\rVert_2^2\leq V(x_t,u^T,y^T).
\end{align*}
\\
\textbf{(c.2). Local upper Bound on $\mathbf{V}$}\\
Let $x_t$ satisfy $\lVert x_t-x^{sr}\rVert_2\leq\delta$ for a sufficiently small $\delta>0$. Since $(u^{sr},y^{sr})\in\mathbb{U}^s\times\mathbb{Y}^s\subseteq\text{int}(\mathbb{U}\times\mathbb{Y})$, $L\geq2n$, and by controllability, there exists a feasible input-output trajectory steering the state to $x^{sr}$ in $n\leq L-n$ steps, while satisfying
\begin{align}\label{eq:thm_proof_ctrb_bound}
\left\lVert\begin{bmatrix}\bar{u}_{[0,\JB{L}]}(t)-u_{\JB{L+1}}^{sr}\\\bar{y}_{[0,\JB{L}]}(t)-y_{\JB{L+1}}^{sr}
\end{bmatrix}\right\rVert_2^2\leq
\Gamma_{uy}\lVert x_t-x^{sr}\rVert_2^2
\end{align}
for some $\Gamma_{uy}>0$.
For the artificial equilibrium, we consider the candidate solution $u^s(t)=u^{sr},y^s(t)=y^{sr}$.
Finally, by Theorem~\ref{thm:traj_rep}, there exists an $\alpha(t)$ satisfying~\eqref{eq:tracking_MPC1}, which implies that the defined trajectory satisfies all constraints of~\eqref{eq:tracking_MPC}.
Hence, a local quadratic upper bound on $V$ can be obtained as
\begin{align*}
V(x_t,u^T,y^T)\leq (\Gamma_{uy}\lambda_{\max}(Q,R)+\gamma\lambda_{\max}(P))\lVert x_t-x^{sr}\rVert_2^2.
\end{align*}
\textbf{(c.3). Exponential Decay of $\mathbf{V}$}\\
Define $n$ candidate solutions for $i\in\mathbb{I}_{[1,n]}$, similar to Part (a) of the proof, as
\begin{align*}
&\bar{u}_k'(t+i)=\bar{u}_{k+1}^*(t+i-1),\>\>\bar{y}_k'(t+i)=\bar{y}_{k+1}^*(t+i-1),
\end{align*}
for $k\in\mathbb{I}_{[-n,\JB{L-1}]}$, and $\bar{u}_{\JB{L}}'(t+i)=u^s\text{$'$}(t+i),\bar{y}_{\JB{L}}'(t+i)=y^s\text{$'$}(t+i)$.
The candidate artificial equilibria are defined as
\begin{align*}
&u^s\text{$'$}(t+i)=u^{s*}(t+i-1),\>\>y^s\text{$'$}(t+i)=y^{s*}(t+i-1),
\end{align*}
and $\alpha'(t+i)$ as a corresponding solution to~\eqref{eq:thm_hankel}.
Using this candidate solution, it is readily shown that, for any $i\in\mathbb{I}_{[1,n]}$, the optimal cost is non-increasing with
\begin{align}\nonumber
&J_L^*(x_{t+i},u^T,y^T)\\\label{eq:cost_iterative_decay}
\leq &J_L(x_{t+i},u^T,y^T,u^s\text{$'$}(t+i),y^s\text{$'$}(t+i),\alpha'(t+i))\\\nonumber
=&J_L^*(x_{t+i-1},u^T,y^T)-\lVert u_{t+i-1}-u^{s*}(t+i-1)\rVert_R^2\\\nonumber
&-\lVert y_{t+i-1}-y^{s*}(t+i-1)\rVert_Q^2.
\end{align}
We derive now a decay-bound of $J_L^*$ over $n$ steps by studying different cases.\\
\textbf{Case 1:}
Assume
\begin{align}\label{eq:thm_nom_case1}
&\sum_{i=0}^{n-1}\lVert u_{t+i}-u^{s*}(t+i)\rVert_R^2+\lVert y_{t+i}-y^{s*}(t+i)\rVert_Q^2\\\nonumber
&\geq\gamma_1\sum_{i=0}^{n-1}\lVert u^{s*}(t+i)-u^{sr}\rVert_S^2+\lVert y^{s*}(t+i)-y^{sr}\rVert_T^2,
\end{align}
for a constant $\gamma_1>0$, which will be fixed later in the proof.
It follows from~\eqref{eq:cost_iterative_decay} that the optimal cost over $n$ steps decreases as
\begin{align}\label{eq:thm_nom_cost_decay1}
&J_L^*(x_{t+n},u^T,y^T)-J_L^*(x_t,u^T,y^T)\\\nonumber
&=\sum_{i=0}^{n-1}J_L^*(x_{t+i+1},u^T,y^T)-J_L^*(x_{t+i},u^T,y^T)\\\nonumber
&\leq-\sum_{i=0}^{n-1}\left(\lVert u_{t+i}-u^{s*}(t+i)\rVert_R^2+\lVert y_{t+i}-y^{s*}(t+i)\rVert_Q^2\right),
\end{align}
where a telescoping sum argument is used for the first equality.
Using $a^2+b^2\geq\frac{1}{2}(a+b)^2$,~\eqref{eq:thm_nom_cost_decay1} implies
\begin{align}\label{eq:thm_nom_proof_decay1}
&J_L^*(x_{t+n},u^T,y^T)-J_L^*(x_t,u^T,y^T)\\\nonumber
&\stackrel{\eqref{eq:thm_nom_case1}}{\leq}-\frac{\gamma_1}{2}\sum_{i=0}^{n-1}\left(\lVert u^{s*}(t+i)-u^{sr}\rVert_S^2+\lVert y^{s*}(t+i)-y^{sr}\rVert_T^2\right)\\\nonumber
&-\frac{1}{2}\sum_{i=0}^{n-1}\left(\lVert u_{t+i}-u^{s*}(t+i)\rVert_R^2+\lVert y_{t+i}-y^{s*}(t+i)\rVert_Q^2\right)\\\nonumber
&\leq-c_3\frac{\min\{1,\gamma_1\}}{4}\sum_{i=0}^{n-1}\left(\lVert u_{t+i}-u^{sr}\rVert_2^2+\lVert y_{t+i}-y^{sr}\rVert_2^2\right),
\end{align}
where $c_3=\lambda_{\min}(Q,R,S,T)$.\\
\textbf{Case 2:} Assume
\begin{align}\label{eq:thm_nom_case2}
&\sum_{i=0}^{n-1}\lVert u_{t+i}-u^{s*}(t+i)\rVert_R^2+\lVert y_{t+i}-y^{s*}(t+i)\rVert_Q^2\\\nonumber
&\leq\gamma_1\sum_{i=0}^{n-1}\lVert u^{s*}(t+i)-u^{sr}\rVert_S^2+\lVert y^{s*}(t+i)-y^{sr}\rVert_T^2.
\end{align}
\textbf{Case 2a:} Assume further
\begin{align}\label{eq:thm_nom_case2a}
\sum_{i=0}^{n-1}\lVert x_{t+i}-x^{s*}(t+i)\rVert_2^2\leq\gamma_2\sum_{i=0}^{n-1}\lVert x^{s*}(t+i)-x^{sr}\rVert_2^2,
\end{align}
for a constant $\gamma_2>0$, which will be fixed later in the proof.
We consider now a different candidate solution at time $t+1$ with artificial equilibrium $\hat{u}^{s}(t+1)=\lambda u^{s*}(t)+(1-\lambda)u^{sr}$, $\hat{y}^{s}(t+1)=\lambda y^{s*}(t)+(1-\lambda)y^{sr}$ for some $\lambda\in(0,1)$, which will be fixed later in the proof.
Clearly, this is a feasible equilibrium and it holds for the corresponding equilibrium state that $\hat{x}^{s}(t+1)=\lambda x^{s*}(t)+(1-\lambda)x^{sr}$.
Further,
\begin{align}\label{eq:thm_nom_case2a_eq1}
\hat{x}^{s}(t+1)-x^{s*}(t)=(1-\lambda)(x^{sr}-x^{s*}(t)),
\end{align}
and similarly for the input and output.
Due to compactness of $\mathbb{U},\mathbb{Y}$, the right-hand side of~\eqref{eq:thm_nom_case2a} is bounded from above by $\gamma_2 x_{\max}$ for some $x_{\max}>0$.
Thus, if $\gamma_2$ is sufficiently small, then $\sum_{i=0}^{n-1}\lVert x_{t+i}-x^{s*}(t+i)\rVert_2^2$ is arbitrarily small as well.
Hence, if in addition $(1-\lambda)$ is sufficiently small, then, by controllability and since $(\hat{u}^{s}(t+1),\hat{y}^{s}(t+1))\in\text{int}(\mathbb{U}\times\mathbb{Y})$, there exists a feasible input-output trajectory $\hat{u}(t+1),\hat{y}(t+1)$ steering the system to $(\hat{u}^s(t+1),\hat{y}^s(t+1))$ in $n$ steps.
Moreover, there exists a corresponding $\hat{\alpha}(t+1)$ satisfying all constraints of~\eqref{eq:tracking_MPC}.
Further, it holds that
\begin{align*}
&\lVert \hat{u}^{s}(t+1)-u^T\rVert_S^2-\lVert u^{s*}(t)-u^T\rVert_S^2\\
=&\left(\hat{u}^{s}(t+1)-u^{s*}(t)\right)^\top S\left(\hat{u}^{s}(t+1)+u^{s*}(t)-2u^T\right)\\
=&(1-\lambda)\left(u^{sr}-u^{s*}(t)\right)^\top S\\
&\cdot\left((1+\lambda)u^{s*}(t)+(1-\lambda)u^{sr}-2u^T\right)\\
=&-(1-\lambda^2)\lVert u^{s*}(t)-u^{sr}\rVert_S^2\\
&-2(1-\lambda)(u^{s*}(t)-u^{sr})^\top S(u^{sr}-u^T)\\
\leq&-(1-\lambda^2)\lVert u^{s*}(t)-u^{sr}\rVert_S^2,
\end{align*}
where the last inequality follows from noting that $2S(u^{sr}-u^T)$ is the gradient of $\lVert u-u^T\rVert_S^2$ evaluated at $u^{sr}$, and the directional derivative of this function towards any other feasible direction increases, due to convexity of~\eqref{eq:opt_reach_eq} by Assumption~\ref{ass:eq_convex} (compare~\cite{Koehler19b} for details).
Similarly,
\begin{align*}
\lVert \hat{y}^{s}(t+1)-y^T&\rVert_T^2-\lVert y^{s*}(t)-y^T\rVert_T^2\\
&\leq-(1-\lambda^2)\lVert y^{s*}(t)-y^{sr}\rVert_T^2.
\end{align*}
By controllability, there exists $\Gamma_{uy}>0$ as in~\eqref{eq:thm_proof_ctrb_bound} such that
\begin{align*}
&\sum_{k=0}^{\JB{L}}\lVert\hat{u}_k(t+1)-\hat{u}^s(t+1)\rVert_R^2+\lVert\hat{y}_k(t+1)-\hat{y}^s(t+1)\rVert_Q^2\\
&\leq \Gamma_{uy}\lambda_{\max}(Q,R)\lVert x_{t+1}-\hat{x}^s(t+1)\rVert_2^2\\
&\leq2\Gamma_{uy}\lambda_{\max}(Q,R)(\lVert x_{t+1}-x^{s*}(t)\rVert_2^2\\
&\qquad\qquad\qquad\qquad\qquad+\lVert x^{s*}(t)-\hat{x}^s(t+1)\rVert_2^2),
\end{align*}
using again the fact that $(a+b)^2\leq2a^2+2b^2$.
The first term can be bounded as
\begin{align*}
\lVert x_{t+1}-x^{s*}(t)\rVert_2^2\leq&\lVert A\rVert_2^2\lVert x_t-x^{s*}(t)\rVert_2^2\\
&+\lVert B\rVert_2^2\lVert u_t-u^{s*}(t)\rVert_2^2\\
\leq &\underbrace{(\lVert A\rVert_2^2+\lVert B\rVert_2^2\Gamma_{uy})}_{c_4\coloneqq}\lVert x_t-x^{s*}(t)\rVert_2^2,
\end{align*}
where the last inequality follows again from a controllability argument.
Combining the bounds leads to
\begin{align*}
&J_L^*(x_{t+1},u^T,y^T)-J_L^*(x_t,u^T,y^T)\\
&\leq 2c_4\Gamma_{uy}\lambda_{\max}(Q,R)\lVert x_t-x^{s*}(t)\rVert_2^2\\
&+2\Gamma_{uy}\lambda_{\max}(Q,R)(1-\lambda)^2\lVert x^{sr}-x^{s*}(t)\rVert_2^2\\
&-\lVert u_t-u^{s*}(t)\rVert_R^2-\lVert y_t-y^{s*}(t)\rVert_Q^2\\
&-(1-\lambda^2)(\lVert u^{s*}(t)-u^{sr}\rVert_S^2+\lVert y^{s*}(t)-y^{sr}\rVert_T^2).
\end{align*}
Using a similar candidate solution at time $t+i$ for $i\in\mathbb{I}_{[2,n]}$, the following can be shown.
\begin{align*}
&J_L^*(x_{t+n},u^T,y^T)-J_L^*(x_t,u^T,y^T)\\
&\leq 2c_4\Gamma_{uy}\lambda_{\max}(Q,R)\sum_{i=0}^{n-1}\lVert x_{t+i}-x^{s*}(t+i)\rVert_2^2\\
&+2\Gamma_{uy}\lambda_{\max}(Q,R)(1-\lambda)^2\sum_{i=0}^{n-1}\lVert x^{sr}-x^{s*}(t+i)\rVert_2^2\\
&-\sum_{i=0}^{n-1}(\lVert u_{t+i}-u^{s*}(t+i)\rVert_R^2+\lVert y_{t+i}-y^{s*}(t+i)\rVert_Q^2)\\
&-(1-\lambda^2)\sum_{i=0}^{n-1}(\lVert u^{s*}(t+i)-u^{sr}\rVert_S^2+\lVert y^{s*}(t+i)-y^{sr}\rVert_T^2)\\
&\stackrel{\eqref{eq:bound_xuy_1},\eqref{eq:thm_nom_case2a}}{\leq}
(2c_4\gamma_2+2(1-\lambda)^2)\Gamma_{uy}\lambda_{\max}(Q,R)c_{x,2}\\
&\qquad\quad\cdot\sum_{i=0}^{n-1}(\lVert u^{s*}(t+i)-u^{sr}\rVert_2^2+\lVert y^{s*}(t+i)-y^{sr}\rVert_2^2)\\
&-\sum_{i=0}^{n-1}(\lVert u_{t+i}-u^{s*}(t+i)\rVert_R^2+\lVert y_{t+i}-y^{s*}(t+i)\rVert_Q^2\\
&-(1-\lambda^2)\sum_{i=0}^{n-1}(\lVert u^{s*}(t+i)-u^{sr}\rVert_S^2+\lVert y^{s*}(t+i)-y^{sr}\rVert_T^2)\\
&\leq -c_5\sum_{i=0}^{n-1}(\lVert u_{t+i}-u^{sr}\rVert_2^2+\lVert y_{t+i}-y^{sr}\rVert_2^2),
\end{align*}
for some $c_5>0$, where the last inequality holds if $\gamma_2$ is sufficiently small and $\lambda$ is sufficiently close to $1$.\\
\textbf{Case 2b:} Assume
\begin{align}\label{eq:thm_nom_case2b}
\sum_{i=0}^{n-1}\lVert x_{t+i}-x^{s*}(t+i)\rVert_2^2\geq\gamma_2\sum_{i=0}^{n-1}\lVert x^{s*}(t+i)-x^{sr}\rVert_2^2.
\end{align}
This implies the existence of an index $k\in\mathbb{I}_{[0,n-1]}$ such that
\begin{align}\label{eq:thm_nom_case2bk}
\lVert x_{t+k}-x^{s*}(t+k)\rVert_2^2\geq\frac{\gamma_2}{n}\sum_{i=0}^{n-1}\lVert x^{s*}(t+i)-x^{sr}\rVert_2^2.
\end{align}
The following auxiliary result will be central for the proof of Case 2b.

\begin{lemma}\label{lem:proof}
There exist $\gamma_3>0,j\in\mathbb{I}_{[0,n-1]}$ such that
\begin{align}\label{eq:thm_nom_case2b1}
&\lVert u_{t+j}-u^{s*}(t+k)\rVert_2^2+\lVert y_{t+j}-y^{s*}(t+k)\rVert_2^2\\\nonumber
\geq&\gamma_3\sum_{i=0}^{n-1}\lVert u^{s*}(t+i)-u^{sr}\rVert_2^2+\lVert y^{s*}(t+i)-y^{sr}\rVert_2^2,
\end{align}
with $k$ as in~\eqref{eq:thm_nom_case2bk}.
\end{lemma}
\begin{pf}
Using the system dynamics, it holds that
\begin{align}\label{eq:thm_proof_assertion1}
\lVert x_{t+k}-x^{s*}(t+k)\rVert_2^2&\leq a_1\lVert x_t-x^{s*}(t+k)\rVert_2^2\\
\nonumber
&+a_2\lVert u_{[t,t+k-1]}-u^{s*}_{k}(t+k)\rVert_2^2,
\end{align}
for suitable $a_1,a_2>0$.
Further, for the observability matrix $\Phi$ and a suitable matrix $\Gamma$, which depends on $B,C,D$, we obtain
\begin{align*}
y_{[t,t+n-1]}-y^{s*}_n(t+k)&=\Phi(x_t-x^{s*}(t+k))\\
&+\Gamma(u_{[t,t+n-1]}-u^{s*}_n(t+k)).
\end{align*}
By observability, we can solve the latter equation for $x_t-x^{s*}(t+k)$, which leads, together with~\eqref{eq:thm_proof_assertion1}, to
\begin{align}
&\lVert x_{t+k}-x^{s*}(t+k)\rVert_2^2\\\nonumber
\leq&a_3\sum_{i=0}^{n-1}(\lVert u_{t+i}-u^{s*}(t+k)\rVert_2^2+\lVert y_{t+i}-y^{s*}(t+k)\rVert_2^2),
\end{align}
for a suitable $a_3>0$.
Let $j$ be the index, for which
\begin{align*}
\lVert u_{t+j}-u^{s*}(t+k)\rVert_2^2+\lVert y_{t+j}-y^{s*}(t+k)\rVert_2^2
\end{align*}
is maximal, which implies
\begin{align}\label{eq:thm_nom_case2b2}
&\lVert u_{t+j}-u^{s*}(t+k)\rVert_2^2+\lVert y_{t+j}-y^{s*}(t+k)\rVert_2^2\\\nonumber
\geq&\frac{1}{n}\sum_{i=0}^{n-1}\lVert u_{t+i}-u^{s*}(t+k)\rVert_2^2+\lVert y_{t+i}-y^{s*}(t+k)\rVert_2^2,
\end{align}
which in turn leads to
\begin{align}\label{eq:thm_proof_assertion2}
&\lVert x_{t+k}-x^{s*}(t+k)\rVert_2^2\\\nonumber
\leq&a_3n(\lVert u_{t+j}-u^{s*}(t+k)\rVert_2^2+\lVert y_{t+j}-y^{s*}(t+k)\rVert_2^2).
\end{align}
Combining~\eqref{eq:thm_proof_assertion2} with~\eqref{eq:thm_nom_case2bk} and~\eqref{eq:bound_xuy_2} concludes the proof of Lemma~\ref{lem:proof}. $\hfill\square$
\end{pf}


It follows from~\eqref{eq:thm_nom_case2} and~\eqref{eq:thm_nom_case2b1} that $j\neq k$, as long as $\gamma_1<\gamma_3$, which will be assumed in the following.
At time $t+j$, we define now a different candidate solution as a convex combination between the optimal solution and the candidate solution from Case 1, i.e.,
\begin{align*}
\bar{u}''(t+j)&=\beta\bar{u}'(t+j)+(1-\beta)\bar{u}^*(t+j),
\end{align*}
for some $\beta\in[0,1]$, with $\bar{u}'(t+j)$ as in the beginning of Part (c.3) of the proof.
The other variables $\bar{y}''(t+j),u^s\text{$''$}(t+j),y^s\text{$''$}(t+j),\alpha''(t+j)$ are defined analogously.
By convexity, this is a feasible solution to~\eqref{eq:tracking_MPC}.
Hence,
\begin{align*}
&J_L^*(x_{t+j},u^T,y^T)\\
\leq&J_L(x_{t+j},u^T,y^T,u^s\text{$''$}(t+j),y^s\text{$''$}(t+j),\alpha''(t+j))\\
\leq&\beta J_L(x_{t+j},u^T,y^T,u^s\text{$'$}(t+j),y^s\text{$'$}(t+j),\alpha'(t+j))\\
&+(1-\beta)J_L^*(x_{t+j},u^T,y^T)\\
&-2\bar{c}\beta(1-\beta)\lVert u^{s*}(t+j)-u^{s}\text{$'$}(t+j)\rVert_2^2\\
&-2\bar{c}\beta(1-\beta)\lVert y^{s*}(t+j)-y^{s}\text{$'$}(t+j)\rVert_2^2,\\
\stackrel{\eqref{eq:cost_iterative_decay}}{\leq}&\beta J_L^*(x_{t+j-1},u^T,y^T)+(1-\beta)J_L^*(x_{t+j},u^T,y^T)\\
&-2\bar{c}\beta(1-\beta)\lVert u^{s*}(t+j)-u^{s}\text{$'$}(t+j)\rVert_2^2\\
&-2\bar{c}\beta(1-\beta)\lVert y^{s*}(t+j)-y^{s}\text{$'$}(t+j)\rVert_2^2,
\end{align*}
where the second inequality follows from strong convexity of~\eqref{eq:opt_reach_eq} for some $\bar{c}>0$.
Fixing $\beta=\frac{1}{2}$ and dividing by $\beta$, we obtain
\begin{align*}
&J_L^*(x_{t+j},u^T,y^T)-J_L^*(x_{t+j-1},u^T,y^T)\\
\leq &-\bar{c}\lVert u^{s*}(t+j)-u^{s*}(t+j-1)\rVert_2^2\\
&-\bar{c}\lVert y^{s*}(t+j)-y^{s*}(t+j-1)\rVert_2^2.
\end{align*}
Suppose now $j>k$.
Then, by defining similar candidate solutions at time instances $t+k,\dots,t+j$, and applying $a^2+b^2\geq\frac{1}{2}(a+b)^2$ repeatedly $j-k-1$ times, we obtain
\begin{align*}
&J_L^*(x_{t+j},u^T,y^T)-J_L^*(x_{t+k},u^T,y^T)\leq-\frac{\bar{c}}{2^{j-k-1}}\cdot\\
&\left(\lVert u^{s*}(t+j)-u^{s*}(t+k)\rVert_2^2+\lVert y^{s*}(t+j)-y^{s*}(t+k)\rVert_2^2\right).
\end{align*}
Conversely, if $k>j$, we arrive at
\begin{align*}
&J_L^*(x_{t+k},u^T,y^T)-J_L^*(x_{t+j},u^T,y^T)\leq-\frac{\bar{c}}{2^{k-j-1}}\cdot\\
&\left(\lVert u^{s*}(t+j)-u^{s*}(t+k)\rVert_2^2+\lVert y^{s*}(t+j)-y^{s*}(t+k)\rVert_2^2\right).
\end{align*}
Combining the two cases and noting that $j\neq k$ (cf. Lemma~\ref{lem:proof}), it follows from~\eqref{eq:cost_iterative_decay} that
\begin{align}\nonumber
&J_L^*(x_{t+n},u^T,y^T)-J_L^*(x_{t},u^T,y^T)\\\label{eq:thm_proof_case2b_decay}
\leq&-\frac{\bar{c}}{2^{|j-k|-1}}\lVert u^{s*}(t+j)-u^{s*}(t+k)\rVert_2^2\\\nonumber
&-\frac{\bar{c}}{2^{|j-k|-1}}\lVert y^{s*}(t+j)-y^{s*}(t+k)\rVert_2^2.
\end{align}
Now, we bound the right-hand side of~\eqref{eq:thm_proof_case2b_decay} by using
\begin{subequations}\label{eq:thm_proof_ineq}
\begin{align}\label{eq:thm_proof_ineqa}
a+b&=\sqrt{(a+b)^2}\geq\sqrt{a^2+b^2},\\\label{eq:thm_proof_ineqb}
a+b&=\sqrt{(a+b)^2}\leq\sqrt{2a^2+2b^2},
\end{align}
\end{subequations}
which hold for any $a,b\geq0$.
In the following, let $\gamma_1$ be sufficiently small such that $\gamma_1<\min\{1,\frac{1}{8c_6}\}\gamma_3$, where $c_6=\frac{\lambda_{\max}(S,T)}{\lambda_{\min}(Q,R)}>0$.
Note that this implies $\gamma_1<\gamma_3$ and hence $j\neq k$, as is required above.
Moreover,
\begin{align*}
&\lVert u^{s*}(t+j)-u^{s*}(t+k)\rVert_2+\lVert y^{s*}(t+j)-y^{s*}(t+k)\rVert_2\\
&\geq\lVert u_{t+j}-u^{s*}(t+k)\rVert_2-\lVert u_{t+j}-u^{s*}(t+j)\rVert_2\\
&\quad+\lVert y_{t+j}-y^{s*}(t+k)\rVert_2-\lVert y_{t+j}-y^{s*}(t+j)\rVert_2\\
&\stackrel{\eqref{eq:thm_nom_case2b1},\eqref{eq:thm_proof_ineqa}}{\geq} \frac{1}{2}\lVert u_{t+j}-u^{s*}(t+k)\rVert_2+\frac{1}{2}\lVert y_{t+j}-y^{s*}(t+k)\rVert_2\\
&+\frac{\sqrt{\gamma_3}}{2}\sqrt{\sum_{i=0}^{n-1}\lVert u^{s*}(t+i)-u^{sr}\rVert_2^2+\lVert y^{s*}(t+i)-y^{sr}\rVert_2^2}\\
&-\lVert u_{t+j}-u^{s*}(t+j)\rVert_2-\lVert y_{t+j}-y^{s*}(t+j)\rVert_2\\
&\stackrel{\eqref{eq:thm_nom_case2},\eqref{eq:thm_proof_ineqb}}{\geq}
\frac{1}{2}\lVert u_{t+j}-u^{s*}(t+k)\rVert_2+\frac{1}{2}\lVert y_{t+j}-y^{s*}(t+k)\rVert_2\\
&+\left(\frac{\sqrt{\gamma_3}}{2}-\sqrt{2\gamma_1c_6}\right)\\
&\cdot\sqrt{\sum_{i=0}^{n-1}\lVert u^{s*}(t+i)-u^{sr}\rVert_2^2+\lVert y^{s*}(t+i)-y^{sr}\rVert_2^2}\\
&\stackrel{\eqref{eq:thm_nom_case2b2}}{\geq}\frac{1}{2}\sqrt{\frac{1}{n}\sum_{i=0}^{n-1}\lVert u_{t+i}-u^{s*}(t+k)\rVert_2^2+\lVert y_{t+i}-y^{s*}(t+k)\rVert_2^2}\\
&+\left(\frac{\sqrt{\gamma_3}}{2}-\sqrt{2\gamma_1c_6}\right)\\
&\cdot\sqrt{\lVert u^{s*}(t+k)-u^{sr}\rVert_2^2+\lVert y^{s*}(t+k)-y^{sr}\rVert_2^2}\\
&\geq c_7\sum_{i=0}^{n-1}\lVert u_{t+i}-u^{sr}\rVert_2+\lVert y_{t+i}-y^{sr}\rVert_2,
\end{align*}
for a suitable $c_7>0$, where the last inequality follows from an inequality similar to~\eqref{eq:thm_proof_ineqb}.
This, together with~\eqref{eq:thm_proof_case2b_decay} implies the existence of a constant $c_8>0$ such that
\begin{align*}
&J_L^*(x_{t+n},u^T,y^T)-J_L^*(x_t,u^T,y^T)\\
\leq &-c_8\sum_{i=0}^{n-1}\left(\lVert u_{t+i}-u^{sr}\rVert_2^2+\lVert y_{t+i}-y^{sr}\rVert_2^2\right).
\end{align*}
\textbf{Combination:}
Combining all cases, there exists some $c_9>0$ such that
\begin{align}\label{eq:thm_proof_combination}
&J_L^*(x_{t+n},u^T,y^T)-J_L^*(x_t,u^T,y^T)\\\nonumber
\leq&-c_9\sum_{i=0}^{n-1}\lVert u_{t+i}-u^{sr}\rVert_2^2+\lVert y_{t+i}-y^{sr}\rVert_2^2.
\end{align}
Furthermore, the IOSS Lyapunov function $W$ satisfies
\begin{align}\nonumber
&W(x_{t+1}-x^{sr})-W(x_t-x^{sr})\\\label{eq:thm_nom_IOSS1}
=&W(A(x_t-x^{sr})+B(u_t-u^{sr}))-W(x_t-x^{sr})\\\nonumber
\leq&-\frac{1}{2}\lVert x_t-x^{sr}\rVert_2^2+c_1 \lVert u_t-u^{sr}\rVert_2^2
+c_2\lVert y_t-y^{sr}\rVert_2^2.
\end{align}
Applying this inequality recursively, we arrive at
\begin{align}\nonumber
W(x_{t+n}-&x^{sr})-W(x_t-x^{sr})\\\label{eq:thm_nom_IOSS}
\leq\sum_{i=0}^{n-1}&-\frac{1}{2}\lVert x_{t+i}-x^{sr}\rVert_2^2+c_1\lVert u_{t+i}-u^{sr}\rVert_2^2\\\nonumber
&+c_2\lVert y_{t+i}-y^{sr}\rVert_2^2.
\end{align}
Thus, choosing $\gamma=\frac{c_9}{\max\{c_1,c_2\}}>0$, the bounds~\eqref{eq:thm_proof_combination} and~\eqref{eq:thm_nom_IOSS} can be used to bound the Lyapunov function candidate $V(x_t,u^T,y^T)$ as
\begin{align*}
V(x_{t+n},u^T,y^T)-V(x_t,u^T,y^T)\leq-\frac{\gamma}{2}\sum_{i=0}^{n-1}\lVert x_{t+i}-x^{sr}\rVert_2^2.
\end{align*}
Due to Parts (c.1) and (c.2) of the proof, $V$ is locally quadratically lower and upper bounded.
Thus, the equilibrium $x^{sr}$ is exponentially stable by standard Lyapunov arguments.
$\hfill\square$
\end{pf}

The proof of Theorem~\ref{thm:main} follows the lines of~\cite{Koehler19b}.
The main difference lies in the fact that an $n$-step analysis of the closed loop is performed.
Intuitively, this can be explained by noting that the input-output behavior over $n$ steps allows to draw conclusions on the behavior of the internal state, which is relevant for stability.
Further, by making an additional case distinction in Case 2b of the proof, we show a decay in the Lyapunov function for the scenario that the actual input and output values $(u_{t+i},y_{t+i})$ are close to their respective artificial equilibria $(u^{s*}(t+i),y^{s*}(t+i))$ for $i\in\mathbb{I}_{[0,n-1]}$, but at least one internal state $x_{t+k}$ is not close to $x^{s*}(t+k)$ with $k\in\mathbb{I}_{[0,n-1]}$.
As is shown in Lemma~\ref{lem:proof}, this implies that, along $n$ time steps, at least one input or output must be distant from the artificial equilibrium $(u^{s*}(t+k),y^{s*}(t+k))$.
By using this insight and defining a new candidate solution as a convex combination of a simpler candidate solution and the optimal solution, a suitable decay of the optimal cost can be shown.
Finally, an IOSS Lyapunov function is employed to translate this decay, which is in terms of input-output values, to a decay in the state.

Theorem~\ref{thm:main} requires compact constraints for Case 2a of the proof, which applies a local controllability argument to treat the case that the state is close to the current artificial steady-state.
The proof is readily extendable to non-compact constraints, when considering initial states within some compact sublevel set of the Lyapunov function $V(x_t,u^T,y^T)\leq V_{\max}$, for a given level $V_{\max}$.
The only difference in this scenario is that the size of $\gamma_2$, and hence also the exponential decay rate of $V$, depends on $V_{\max}$.
That is, for larger initial values, the convergence rate derived in the proof decreases.

\section{Numerical example}\label{sec:example}
In this section, we apply the developed tracking MPC scheme to a four tank system, which has been considered in~\cite{raff2006nonlinear}.
This example is well-known as an open-loop stable system which can be rendered unstable by an MPC without terminal constraints if the horizon is chosen too short.
It was also considered by~\cite{Berberich19b} for data-driven MPC with a fixed terminal equality constraint.
In the following, we show that the present tracking MPC scheme with online optimization of an artificial equilibrium admits a significantly larger region of attraction, without requiring knowledge of the equilibrium input corresponding to a desired output target setpoint. 
The linearized dynamics of the system are
\begin{align*}
x_{k+1}&=\begin{bmatrix}0.921&0&0.041&0\\
0&0.918&0&0.033\\
0&0&0.924&0\\0&0&0&0.937\end{bmatrix}x_k+\begin{bmatrix}
0.017&0.001\\0.001&0.023\\0&0.061\\0.072&0
\end{bmatrix}u_k,\\
y_k&=\begin{bmatrix}1&0&0&0\\0&1&0&0
\end{bmatrix}x_k.
\end{align*}
We assume that the system matrices are unknown, but one input-output trajectory $\{u_k^d,y_k^d\}_{k=0}^{N-1}$ of length $N=100$ is available, which is generated by sampling $u_k^d$ uniformly from $[-1,1]^2$.
This trajectory is used via the proposed MPC scheme in order to track the desired target output $y^T=\begin{bmatrix}1&1\end{bmatrix}^\top$, without specification of an input setpoint, i.e., the input weight is set to $S=0$.
Further, we impose constraints on the input and output as $\mathbb{U}=[-1.2,1.2]\times[-2,2],\mathbb{Y}=[0,1.2]^2$.
The equilibrium constraints are chosen as $\mathbb{U}^s=0.99\mathbb{U},\mathbb{Y}^s=0.99\mathbb{Y}$.

\begin{figure}[h!]
\begin{center}
\includegraphics[width=0.5\textwidth]{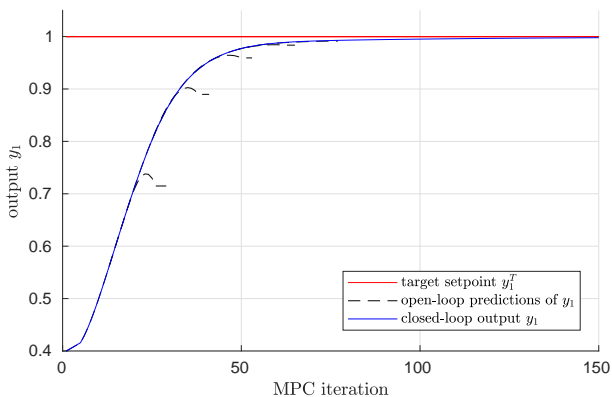}
\end{center}
\caption{First component of the target setpoint $y^T$, of the open-loop predictions $\bar{y}^*(t)$ at multiple time instances $t\in\{0,12,24,36,48\}$, and of the closed-loop trajectory $\{y_t\}_{t=0}^{150}$.}
\label{fig:y1_tracking}
\end{figure}

The prediction horizon is set to $L=\JB{24}$, which is the maximal prediction horizon such that an input trajectory of length $N=100$ can be persistently exciting of order\footnote{Note that, for the matrix $H_{\JB{L+2n+1}}(u^d)$ to have full row rank, it must hold that \JB{$N\geq(m+1)(L+2n+1)-1$}.} \JB{$L+2n+1$}.
Further, the cost matrices are defined as $Q=5I_2,R=I_2,T=200I_2$.
For a given trajectory, there exist infinitely many (arbitrarily large) vectors $\alpha(t)$ satisfying~\eqref{eq:tracking_MPC1} and, therefore, a direct implementation of the proposed scheme can be numerically ill-conditioned.
Therefore, we include a norm-penalty of the form $10^{-4}\cdot\lVert \alpha(t)\rVert_2^2$ in the stage cost, whose utility was thoroughly analyzed in~\cite{Berberich19b} for a robust data-driven MPC scheme in the presence of noise.
Figure~\ref{fig:y1_tracking} illustrates the closed-loop behavior of the first component of the output as well as multiple exemplary open-loop predictions.
It can be seen that the artificial equilibrium $y^{s*}(t)$ is updated continuously and converges to the desired target setpoint, which is in this case equal to the optimal reachable equilibrium, i.e., $y^{sr}=y^T$.
Thus, also the closed-loop output converges to the target setpoint.

Compared to the scheme of~\cite{Berberich19b}, which relied only on terminal equality constraints, but not on an online optimization of the setpoint, the present scheme exhibits several advantages.
First of all, to apply the scheme of~\cite{Berberich19b}, the optimal reachable equilibrium input $u^{sr}=\begin{bmatrix}1&1.8\end{bmatrix}^\top$ needs to be computed explicitly, which is non-trivial without model knowledge, whereas the present scheme computes $u^{sr}$ automatically.
Further, optimizing the setpoint online increases the size of the region of attraction, since the terminal equality constraints do not need to be satisfied already in the first iteration.
In particular, for the above example, the scheme of~\cite{Berberich19b} is only initially feasible for \emph{significantly} larger prediction horizons.
This requires a) more computational power and b) a significantly longer data trajectory.
Furthermore, the present scheme leads to smoother closed-loop trajectories with less overshoot, compared to the scheme without artificial equilibrium.

\section{Conclusion}\label{sec:conclusion}
We presented a novel data-driven tracking MPC scheme which can cope with unreachable setpoints, which may potentially change online.
The scheme is purely data-driven and does not require any model knowledge.
Various desirable properties of the closed loop were proven, and the practical applicability of the scheme was illustrated for a realistic example.
Our results are also an important extension of model-based tracking MPC to the case of positive semidefinite stage costs, which could be dealt with in the present paper by showing a cost decay over $n$ consecutive time steps.
Although the presented scheme is in most scenarios more robust than the one with simple terminal equality constraints used in~\cite{Berberich19b}, an important issue for future research is to give robust stability \emph{guarantees} for the proposed tracking MPC scheme in the case of measurement noise.
\bibliographystyle{ifacconf}   
\bibliography{Literature_JK19_different}  

\end{document}